# An Adaptive Differential Privacy Method Based on Federated Learning

Zhiqiang Wang[1,2,*], Xinyue Yu[1], Qianli Huang[1] and Yongguang Gong[1]

[1] Beijing Electronic Science and Technology Institute, Beijing, China
[2] State Information Center, Beijing, China
wangzq@besti.edu.cn

**Abstract.** Differential privacy is one of the methods to solve the problem of privacy protection in federated learning. Setting the same privacy budget for each round will result in reduced accuracy in training. The existing methods of the adjustment of privacy budget consider fewer influencing factors and tend to ignore the boundaries, resulting in unreasonable privacy budgets. Therefore, we proposed an adaptive differential privacy method based on federated learning. The method sets the adjustment coefficient and scoring function according to accuracy, loss, training rounds, and the number of datasets and clients. And the privacy budget is adjusted based on them. Then the local model update is processed according to the scaling factor and the noise. Finally, the server aggregates the noised local model update and distributes the noised global model. The range of parameters and the privacy of the method are analyzed. Through the experimental evaluation, it can reduce the privacy budget by about 16%, while the accuracy remains roughly the same.

## 1  Introduction

In response to industry demands and the challenge of "data island", Google introduced the concept of "federated learning" in 2016 [1]. Federated learning is a distributed architecture [2] that requires multiple clients to locally analyze and compute their respective stored data. The computed results are shared with a global model, without sharing the participants' raw data. This approach makes the data "accessible but not visible", reducing privacy risks and communication complexity associated with traditional machine learning methods, and alleviating the problem of data island. Although federated learning does not involve the direct exchange of raw data, it provides stronger privacy protection compared to traditional machine learning. However, since the computed results themselves contain certain privacy information, there is still a risk of privacy leakage if comprehensive and sufficient privacy safeguards are not in place. Therefore, ensuring privacy protection in federated learning has become one of the research directions in this field.

In the current schemes, differential privacy [3] is one common approach of privacy protection. However, the conventional method with a fixed privacy budget typically ensures security but fails to guarantee the accuracy of the model. Furthermore, when considering adaptive noise addition, the boundaries of privacy budgets are often overlooked, resulting in a smaller or larger assigned privacy budget that makes training less convergent or less secure. Therefore, we proposed an adaptive differential privacy method based on federated learning (cosAFed). This method computes the adjustment coefficient and scoring function before data transmission in federated learning. It then changes the privacy budget based on these results. Then the scaling factor is calculated, noise is generated and the local model updates are processed. Finally, the server aggregates and distributes the noised global model. Our method can enhance data protection while ensuring accuracy. The innovations and contributions of this paper are as follows:

(1) We proposed an algorithm for adjusting the privacy budget. Firstly, this algorithm calculates the adjustment coefficient based on the cosine similarity between the local model and the global model from the previous round, the number of datasets, the number of client datasets, the total number of clients, and the number of selected clients. Then, a scoring function is designed that considers accuracy, loss, and training rounds. Finally, it adjusts the privacy budget based on the adjustment coefficient and scoring results.

(2) We proposed an adaptive differential privacy method based on federated learning. After obtaining the adjusted privacy budget through the algorithm mentioned above, this method first calculates the scaling factor and generates noise, then processes the local model update and adds noise before uploading it, and finally aggregates and distributes the noised global by the server.

(3) We analyzed and evaluated an adaptive differential privacy method based on federated learning. Firstly, the parameter range is analyzed. Then, it satisfies $\varepsilon'$-differential privacy. Finally, compared with LAPFed, ADPFL, and cosFed, it has certain advantages in data protection ability and model accuracy.

## 2  Related Work

Differential privacy is one of the commonly used privacy protection techniques in federated learning. Since 2017, both domestic and international researchers have proposed various solutions in this field. In 2017, Geyer et al. [4] introduced user-level differential privacy protection algorithms specifically for federated learning. They averaged the client models and utilized a randomization mechanism to approximate the average, thereby concealing the contributions of individual clients and protecting their entire datasets. However, the performance of their approach was heavily influenced by the number of clients. It failed to guarantee accuracy or convergence when the number of clients was small. In 2018, McMahan et al. [5] further improved user-level privacy protection by implementing long short-term memory. Nevertheless, their algorithm required pre-designed scaling factors, which meant that the norms of each vector needed to be estimated in advance, posing practical challenges. In 2019, Triastcyn et al. [6] introduced Bayesian differential privacy into federated learning and proposed DP-FedAvg, which provided differential privacy guarantees. In 2020, Liu et al. [7] presented the Adaptive Privacy-preserving Federated Learning (APFL) framework. Based on localized differential privacy, APFL first employed a layered correlation propagation algorithm to calculate the contributions of each attribute class to the output. Then, adaptive noise was injected into the data attributes to mitigate the impact of noise on the results. Additionally, random privacy protection adjustment techniques were introduced to ensure accuracy. However, there were additional costs associated with the server-side pre-training process, client-side computations before training, and interference contributions. In 2021, Yang et al. [8] proposed an effective model perturbation method in federated learning. They introduced a large positive real number as random noise during the transmission of model parameters from clients to the server to perturb the transmitted data，and at the same time ensured that the server obtained the true data by eliminating the additional noise, thereby maintaining accuracy. This method defended against reconstruction and membership inference attacks launched by curious clients. However, its premise required the server to be a secure and trusted third party; otherwise, the global model stored on the server would be in plaintext, posing security risks. In 2022, Ouadrhiri et al. [9] proposed a two-layer privacy protection method. In the first layer, they reduced the dimensionality of the training dataset based on Hensel's lemma, ensuring a reduction in dataset dimensionality without losing information. The second layer applied differential privacy protection to the compressed dataset generated by the first layer. By applying differential privacy once before training, this method overcame privacy leakage issues caused by synthesis. Experimental results showed that the method achieved good accuracy while protecting user privacy, but the accuracy was dependent on dataset compression and privacy budget. In 2022, Chen et al. [10] introduced a gradient compression framework based on federated learning with adaptive privacy budget allocation. In this framework, Top-k compression was applied to the client side for data transmission. The compressed data was then perturbed before transmission, and the server dynamically adjusted the allocated privacy budget based on communication rounds. This framework balanced security, accuracy, and communication efficiency. However, based on the privacy budget allocation method, as the communication rounds increased, the privacy budget decreased, and the added noise increased, which could result in non-convergence after a certain point. In 2022, Zhang [11] proposed an adaptive privacy budget strategy for federated learning based on localized differential privacy. The privacy budget was adjusted according to the cosine similarity, reducing the privacy budget during the training process to protect the models in larger training rounds. However, this approach lacked consideration for factors such as accuracy and loss, and the aggregated results were not evaluated. In 2023, Cao et al. [12] proposed a differential privacy federated learning algorithm based on functional mechanisms. They replaced the perturbed gradient with a perturbed objective function to ensure the availability of transmitted gradients. However, the gradients contained certain privacy information that was not protected during transmission, leading to privacy leakage issues. In 2023, Tang [13] presented an adaptive differential privacy federated learning algorithm. They utilized a scoring function as an evaluation criterion for reducing noise, but the setting of adjustment coefficients was not discussed. In 2023, Wang et al. [14] proposed an Adaptive Differential Privacy Federated Learning framework based on adaptive clipping. They adaptively selected the gradient clipping threshold based on the L2 norm to stabilize the training process. Additionally, Gaussian noise was added to the local updated model parameters before transmission to prevent privacy leakage in federated learning. However, their approach of uniformly allocating privacy budgets resulted in unnecessary waste.

In summary, existing methods usually have a fixed privacy budget, and affect the accuracy. When considering noise adaptive addition, the boundary and influencing factors of the privacy budget are often ignored, resulting in unreasonable adjusts.

## 3 Methodology

### 3.1 Methodology overview

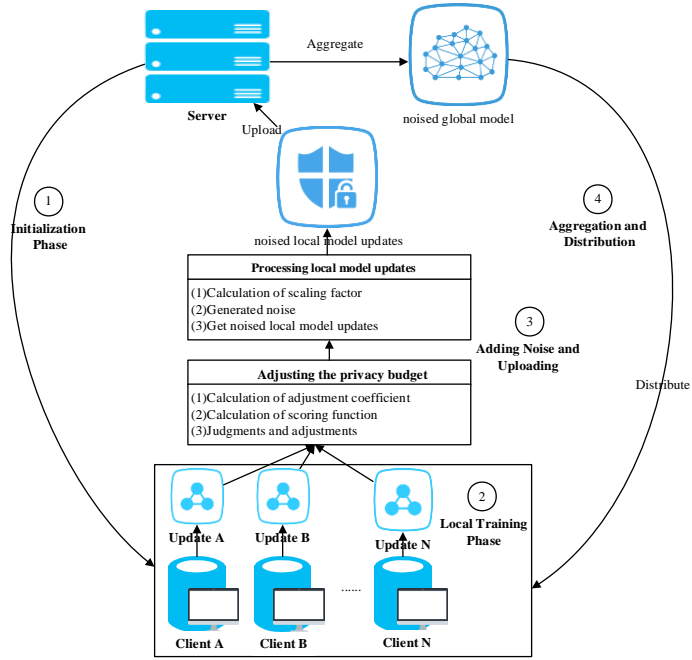

**Fig. 1.** The specific process of this method

An adaptive differential privacy method based on federated learning can be divided into four stages: initialization, local training, adding noise and uploading, and aggregation and distribution. The initialization and local training stages are consistent with the FedAvg [15]. The adding noise and uploading stage involve three parts: adjusting the privacy budget, processing local model updates, and uploading. In the aggregation and distribution stage, the global model is aggregated and distributed, as well as the accuracy and loss are recorded, which are used to adjust the privacy budget in the next round. The specific process of this method is illustrated in Figure 1 and can be represented by Algorithm 1.

---

**Algorithm 1: cosAFed**

---

**Input:** epochs $T$, initial global model $A^0$, clients $C = \{C_1, C_2, \ldots, C_M\}$, number of clients that participate in training $N$, Datasets $D = \{D_1, D_2, \ldots, D_M\}$, learning rate $\eta$, privacy budget $\varepsilon$, sensitivity $\Delta f$

**Output:** the noised global model $A^T$

01  negotiation model $B_1^0, B_2^0, \ldots, B_N^0 = A^0$         /* Initialization */
02  **For** $t \leftarrow 1$ **to** $T$ **do**                     /* Local Training */
03   **For** $i \leftarrow 1$ **to** $N$ **do**
04    $B_i^t \leftarrow SGD(A_i^{t-1}, \eta, D_i)$
05    $\Delta B_i^t \leftarrow A_i^{t-1} - B_i^t$
06    $\varepsilon_i^t \leftarrow ADJUST(M, N, D_i, D, B_i^t, A_i^{t-1}, t, T, acc, loss)$ //Adjusting
07    $\partial \leftarrow \frac{A_i^{t-1}.DOT(B_i^t)}{NP.LINALG.NORM(A_i^{t-1}) * NP.LINALG.NORM(B_i^t)}$  // calculation of scaling factor
08    $noisy \leftarrow N(0, \Delta f / \varepsilon')$    // generated noise
09    $Q_i^t \leftarrow \partial * \Delta B_i^t + noisy$ **and** upload $Q_i^t$
10   **EndFor**
11   $A^t \leftarrow FEDAVG(Q^t, A^{t-1})$ //$Q^t = \{Q_1^t, \ldots, Q_N^t\}$ // Sever Aggregation
12   **SAVE** accuracy **and** loss
13   **SELECT** next clients and **DISTRIBUTE** $A^t$
14  **EndFor**
15  **Return** $A^T$

---

## 3.2 Adjusting the privacy budget

In the adjusting the privacy budget part, we proposed an algorithm to adjust the privacy budget, which involves three steps: calculation of adjustment coefficient, calculation of scoring function, and judgments and adjustments. The detailed algorithm is presented in Algorithm 2.

---

**Algorithm 2: $ADJUST()$**

---

**Input:** epochs $T$, the current round $t$, global model for round $t-1$ $A_i^{t-1}$, the local model of the $i$ client in round $t$ $B_i^t$, the local model update of the $i$ client in round $t$ $\Delta B_i^t$, clients $C = \{C_1, C_2, \ldots, C_M\}$, number of clients that participate in training $N$, privacy budget $\varepsilon$, datasets $D = \{D_1, D_2, \ldots, D_M\}$, $acc = \{acc_1, acc_2, \ldots, acc_{t-1}\}$, $loss = \{loss_1, loss_2, \ldots, loss_{t-1}\}$

**Output:** the adjusted privacy budget $\varepsilon'$

01  $Sc \leftarrow \frac{A_i^{t-1}.DOT(B_i^t)}{NP.LINALG.NORM(A_i^{t-1}) * NP.LINALG.NORM(B_i^t)}$  /* **scaling factor** */
02  **If** $Sc \geq 0$
03   $p \leftarrow ABS(1 - (Sc * M * LEN(D_i))/(N * LEN(D)))$
04  **Else**

```
05  p ← 1
06  If loss_{t-1} ≥ loss_{t-2}     /* Calculation of scoring function */
07    socre_{loss} ← 1
08  Else
09    socre_{loss} ← 0
10  For o ← 0 to N do    //acc
11    If t - 1 - N + o < 0
12      temp_{acc} ← temp_{acc}
13    Else
14      temp_{acc} ← temp_{acc} + acc_{t-1-N+o}
15  EndFor
16  If temp_{acc} ≥ acc_{t-1}
17    socre_{acc} ← 1
18  Else
19    socre_{acc} ← 0
20  If t/T ≥ 1/2
21    socre_t ← 1
22  Else
23    socre_t ← 2 * t/T
24  socre ← 30socre_{loss} + 40socre_{acc} + 30socre_t
25  If socre > 50 and p ≤ 1 do   /* Judgements and adjustments */
26    ε' ← pε,
27  Else do
28    ε' ← ε
```

The adjustment coefficient is influenced by the cosine similarity between the global model and local models, the number of client datasets, the total number of datasets, the number of selected clients, and the total number of clients. When the cosine similarity is non-negative, it indicates that the similarity between the two models is quite high. The higher the similarity, the more noise needs to be introduced to ensure data security. So, in this case, modifying the adjustment coefficient makes the privacy budget smaller. Otherwise, data availability should be guaranteed to ensure accuracy. Therefore, the adjustment coefficient is set to 1, leaving the initial privacy budget unchanged. The calculation is shown in Formula 1, where $p$ represents the adjustment coefficient, $A$ denotes the global model from the previous round, $B$ represents the local model in the current round, $M$ represents the total number of clients in this training, $N$ represents the number of clients selected in the current round, $D_i$ represents the number of datasets for the client $i$, and $D$ represents the total number of datasets.

$$p = \begin{cases} 1 & \cos(A,B) < 0 \\ \left|1 - \cos(A,B) \times \frac{M}{N} \times \frac{D_i}{D}\right| & \cos(A,B) \geq 0 \end{cases} \tag{1}$$

The scoring function is influenced by the training rounds, accuracy, and loss. The scoring function is shown in Formula 2, where $socre$ represents the overall scoring result, $socre_{loss}$ represents the scoring result based on the loss, $socre_{acc}$ represents the scoring result based on the accuracy, and $socre_t$ represents the scoring result based on the training rounds.

$$socre = 30 socre_{loss} + 40 socre_{acc} + 30 socre_t \qquad (2)$$

The loss represents the discrepancy between the predicted output and the actual value [16]. During the training process, it should gradually decrease and eventually stabilize within a small range of fluctuations. The scoring formula for the loss is defined as shown in Formula 3, where $loss_i$ represents the loss in the $i$-th round and $loss_{i-1}$ represents the loss in the $i-1$-th round.

$$socre_{loss} = \begin{cases} 1 & loss_i \geq loss_{i-1} \\ 0 & loss_i < loss_{i-1} \end{cases} \qquad (3)$$

The accuracy should exhibit an increasing trend during the training process and eventually stabilize at a relatively steady state. Since the accuracy fluctuates, the accuracy of all adjacent rounds may vary greatly. Therefore, this algorithm calculates the average accuracy for comparison and evaluation, and the scoring formula for accuracy is defined as shown in Formula 4, where $acc_j$ represents the accuracy in the $j$-th round, $i$ represents the previous round, $N$ represents the number of selected clients in the current training round. For example, if three clients are selected for the current training round, the average accuracy calculation in the 10th round would be $(acc_9 + acc_8 + acc_7)/3$.

$$socre_{acc} = \begin{cases} 1 & \sum_{j=i-N+1}^{i} acc_j /N \geq acc_i \\ 0 & \sum_{j=i-N+1}^{i} acc_j /N < acc_i \end{cases} \qquad (4)$$

In normal circumstances, as the number of training rounds increases, the accuracy tends to improve while the loss decreases. Therefore, the scoring formula for the training rounds is defined as shown in Formula 5, where $t$ represents the current training round and $T$ denotes the total number of training rounds.

$$socre_t = 2 min\left(\frac{t}{T}, \frac{1}{2}\right) \qquad (5)$$

When the $socre$ is less than or equal to 50, the maximum probability is that the number of training rounds is small, the accuracy changes greatly, and the loss becomes smaller, which is the early stage of training. In this case, the local model update changes greatly each round, does not need too strong protection, and the larger privacy budget can promote the accuracy to increase significantly. Cases where $p > 1$ may be due to the small percentage of selected clients compared to the total number of clients. The fewer clients involved in the training, the less accurate the model. Therefore, a large privacy budget is required to keep data as available as possible. Otherwise, when the $socre$ is greater than 50, the local model update changes less and it is easier to infer the original data. In this case, the noise is increased appropriately to protect the data and preserve the original information. Therefore, multiplying the adjustment coefficient by the privacy budget reduces the privacy budget and enhances data protection. The judgments and adjustments are illustrated in Formula 6. In the equation, $\varepsilon'$ represents the adjusted privacy budget, $\varepsilon$ represents the initial privacy budget, $p$ denotes the adjustment coefficient, and $socre$ represents the scoring result.

$$\varepsilon' = \begin{cases} p\varepsilon & socre > 50 \text{ and } p \leq 1 \\ \varepsilon & socre \leq 50 \text{ or } p > 1 \end{cases} \qquad (6)$$

### 3.3 Process of local model updates

The process of local model updates can be divided into three steps: calculation of scaling factor, generated noise, and get noised local model updates. The cosine similarity is used as the criterion to compute the scaling factor. The calculation formula is presented as Formula 7, where $A$ represents the previous round's global model, $B$ represents the local model corresponding to the current client in the current round, $G$ represents the local model update, $G'$ represents the transmitted local model update, $\varepsilon'$ denotes the adjusted privacy budget, $\Delta f$ represents the sensitivity, and $Lap(0, \Delta f/\varepsilon')$ represents the Laplace-distributed noise.

$$G' = cos(A,B)\, G + Lap(0, \Delta f/\varepsilon') \qquad (7)$$

## 4  Bound And Privacy Analysis

### 4.1  Bound analysis

#### 4.1.1 Adjustment coefficients

According to Formula 1, An exploration of its limits is as follows: When the cosine value is less than 0, the adjustment coefficient of 1 is a constant, and a large number of experiments have found that the probability of this situation is very small, so the analysis is not carried out [13]. When cosine is greater than or equal to 0, Let $f(A,B) = cos(A,B)$, where $A$ and $B$ are two vectors whose cosine similarity ranges from -1 to 1, also because $cos(A,B) \geq 0$, so $f(A,B) = cos(A,B) \in [0,1]$. Let $g(M,N) = \frac{M}{N}$, where $M$ and $N$ are positive integers and $M \geq N \geq 1$. So $1 \geq \frac{N}{M} \geq \frac{1}{M}$ and $M \geq \frac{M}{N} \geq 1$, namely $g(M,N) = \frac{M}{N} \in [1,M]$. Let $z(D_i, D) = \frac{D_i}{D}$, where $D_i$ and $D$ are positive integers and $D \geq D_i \geq 1$. So $1 \geq \frac{D_i}{D} \geq \frac{1}{D}$, namely $z(D_i, D) = \frac{D_i}{D} \in [\frac{1}{D}, 1]$. $f(A,B), g(M,N)$, and $z(D_i, D)$ are non-negative and independent of each other.

Case 1:
$$f(A,B) \times g(M,N) \times z(D_i, D) \geq 0 \times 1 \times \frac{1}{D}$$
$$f(A,B) \times g(M,N) \times z(D_i, D) \geq 0$$
$$-f(A,B) \times g(M,N) \times z(D_i, D) \leq 0$$
$$1 - f(A,B) \times g(M,N) \times z(D_i, D) \leq 1$$
$$0 \leq |1 - f(A,B) \times g(M,N) \times z(D_i, D)| \leq 1$$

Case 2:
$$f(A,B) \times g(M,N) \times z(D_i, D) \leq 1 \times M \times 1$$
$$f(A,B) \times g(M,N) \times z(D_i, D) \leq M$$
$$-f(A,B) \times g(M,N) \times z(D_i, D) \geq -M$$
$$1 - f(A,B) \times g(M,N) \times z(D_i, D) \geq 1 - M$$

If $1 - f(A,B) \times g(M,N) \times z(D_i, D) \geq 0$: Because $M \geq 1$, $-M \leq -1$, $1 - M \leq 0$, $|1 - M| = M - 1 \geq 0$. $|1 - f(A,B) \times g(M,N) \times z(D_i, D)| \geq |1 - M| \geq 0$. If not: $|1 - f(A,B) \times g(M,N) \times z(D_i, D)| \leq |1 - M| = M - 1$. To sum up, $|1 - f(A,B) \times$

$g(M,N) \times z(D_i, D)| \in [0,1] \cup [0, M-1]$, $\left|1 - \cos(A,B) \times \frac{M}{N} \times \frac{D_i}{D}\right| \in [0, M-1]$, adjustment coefficients $p \in [0, M-1]$.

### 4.1.2 Scoring function

The score of loss and accuracy is either 0 or 1, so the scoring function of training rounds is mainly analyzed. Formula 5 can be rewritten as Formula 8.

$$socre_t = \begin{cases} \frac{2t}{T} & 0 < \frac{t}{T} \leq \frac{1}{2} \\ 1 & \frac{t}{T} > \frac{1}{2} \end{cases} \quad (8)$$

It is a piecewise function, where the T and t are positive integers and $1 \leq t \leq T$. So $\frac{1}{T} \leq \frac{t}{T} \leq 1$, according to $0 < \frac{t}{T} \leq \frac{1}{2}, \frac{1}{T} \leq \frac{t}{T} \leq \frac{1}{2}, \frac{2}{T} \leq \frac{2t}{T} \leq 1$.Therefore, when $0 < \frac{t}{T} \leq \frac{1}{2}$, $\frac{2}{T} \leq socre_t \leq 1$; when $\frac{t}{T} > \frac{1}{2}, \frac{t}{T} > \frac{1}{2}$. As a result, $socre_t \in [\frac{2}{T}, 1]$. $socre_{loss}$, $socre_{acc}$, and $socre_t$ are non-negative and independent of each other. According to Formula 2:

Case 1:
$$socre \geq 30 \times 0 + 40 \times 0 + 30 \times \frac{2}{T} = \frac{60}{T}$$

Case 2:
$$socre \leq 30 \times 1 + 40 \times 1 + 30 \times 1 = 100$$

To sum up, $\frac{60}{T} \leq socre \leq 100$, $socre \in [\frac{60}{T}, 100]$.

### 4.1.3 Privacy budget

When $socre \leq 50$ or $p > 1$, privacy budget stays the same. Otherwise, it needs to be reduced by adjusting coefficients. According to the analysis, $p \in [0, M-1]$ can be obtained, and according to Formula 6, $p \leq 1$ can be obtained. So $p \in [0,1]$, $\varepsilon' = p\varepsilon \in [0, \varepsilon]$. To sum up, privacy budget $\varepsilon' \in [0, \varepsilon]$.

### 4.2   Privacy analysis

**Theorem 1: Assuming an initial privacy budget of $\varepsilon$, the upload local model updates satisfied $\varepsilon'$-differential privacy.**

**Prove or analyze:** It is known that this algorithm is an improvement on Laplace mechanism, where the adjusted privacy budget $\varepsilon' \in [0, \varepsilon]$ Let $D_i$ and $D'_i$ be adjacent data sets, $F$ be the output function, $G(D'_i) = (g_i^1 + \Delta g_i^1, g_i^2 + \Delta g_i^2, \ldots, g_i^n + \Delta g_i^n)$, output set $Y = (y_i^1, y_i^2, \ldots, y_i^n)$, $\Delta f_i^t = max\|G(D_i) - G(D'_i)\|_1 = max\|\sum_{m=1}^n \Delta g_i^m\|$. According to the Laplace noise probability density formula [17], then each round has:

$$\Pr[F(D_i) \in Y] = \prod_{m=1}^n \frac{\varepsilon'^t_i}{2\Delta f_i^t} e^{\frac{-\varepsilon'^t_i \|Y - G(D_i)\|}{\Delta f_i^t}}$$

$$\Pr[F(D'_i) \in Y] = \prod_{m=1}^n \frac{\varepsilon'^t_i}{2\Delta f_i^t} e^{\frac{-\varepsilon'^t_i \|Y - G(D'_i)\|}{\Delta f_i^t}}$$

$$\frac{\Pr[F(D_i) \in Y]}{\Pr[F(D'_i) \in Y]} = e^{\frac{\varepsilon'^t_i \sum_{m=1}^n (-\|y_i^m - g_i^m\| - \|y_i^m - (g_i^m + \Delta g_i^m)\|)}{\Delta f_i^t}} \leq e^{\frac{\varepsilon'^t_i \|G(D_i) - G(D'_i)\|_1}{\Delta f_i^t}} = e^{\varepsilon'^t_i}$$

The definition of differential privacy is satisfied and $\varepsilon' \in [0, \varepsilon]$. So given an initial privacy budget of $\varepsilon$, the upload local model updates satisfied $\varepsilon'$-differential privacy.

## 5 Experimental Evaluation

### 5.1 Experimental environment

The experimental environment is shown in Table 1, simulating the federated learning training process of 1 server and 100 clients. MNIST dataset and Logistics model were selected. The experimental divided into Equal and Random groups according to different partitioning methods of data sets [18]. The accuracy of the model named best_acc, will be analyzed and compared in the experiment.

Table 1. Experimental environment

| NAME | INFORMATION |
| --- | --- |
| Hardware Information | 11th Gen Intel(R) Core(TM) i5-11300H @ 3.10GHz 2.61 GHz |
| Software Information | Windows 11, PyCharm 2021.2.1, Python 3.7, PyTorch 1.7.1, Anaconda, Numpy, Pandas |

### 5.2 Overall Performance

Based on the different numbers of selected clients and the initial privacy budget, we analyzed the experimental results. The specific experimental parameters are shown in Table 2.

Table 2. Experimental parameters

| Group | Number | $\varepsilon$ | $N$ |
| --- | --- | --- | --- |
| Equal | 1-1 | 100 | 3 |
| | 1-2 | 100 | 5 |
| | 1-3 | 100 | 7 |
| | 1-4 | 120 | 3 |
| | 1-5 | 120 | 5 |

|  | 1-6 | 120 | 7 |
|---|---|---|---|
|  | 1-7 | 150 | 3 |
|  | 1-8 | 150 | 5 |
|  | 1-9 | 150 | 7 |
|  | 1-10 | 170 | 3 |
|  | 1-11 | 170 | 5 |
|  | 1-12 | 170 | 7 |
|  | 1-13 | 200 | 3 |
|  | 1-14 | 200 | 5 |
|  | 1-15 | 200 | 7 |
| Random | 2-1 | 100 | 3 |
|  | 2-2 | 100 | 5 |
|  | 2-3 | 100 | 7 |
|  | 2-4 | 120 | 3 |
|  | 2-5 | 120 | 5 |
|  | 2-6 | 120 | 7 |
|  | 2-7 | 150 | 3 |
|  | 2-8 | 150 | 5 |
|  | 2-9 | 150 | 7 |
|  | 2-10 | 170 | 3 |
|  | 2-11 | 170 | 5 |
|  | 2-12 | 170 | 7 |
|  | 2-13 | 200 | 3 |
|  | 2-14 | 200 | 5 |
|  | 2-15 | 200 | 7 |

### 5.2.1 Accuracy.

Based on the same initial privacy budget and with a different number of selected clients, the training curve results for the Equal group are shown in Figure 2, while the training curve results for the Random group are displayed in Figure 3.

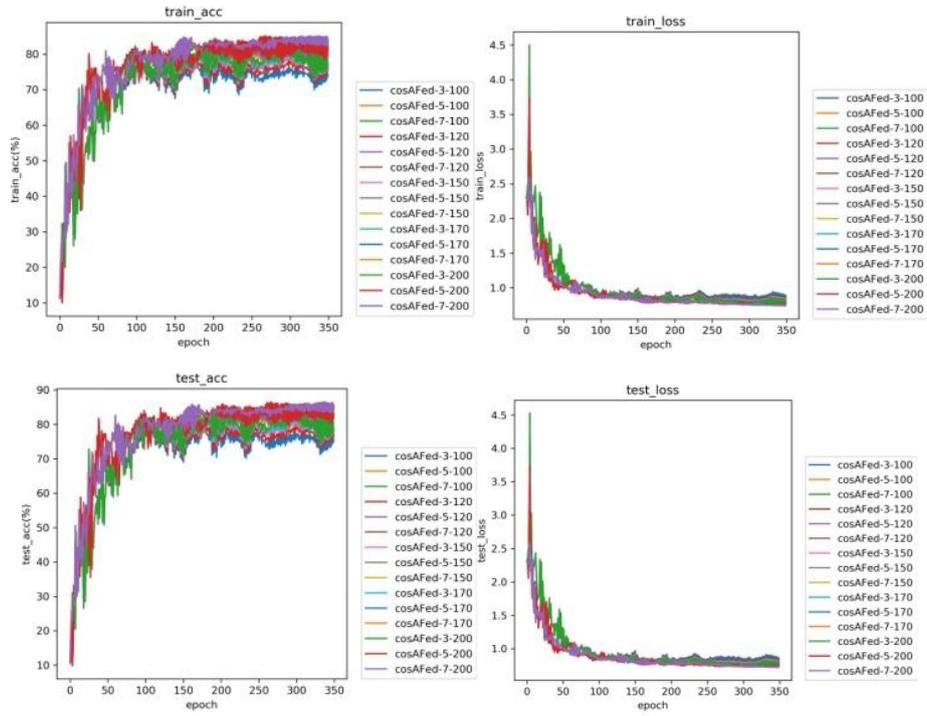

**Fig. 2.** The training results for the Equal group with different number of selected clients

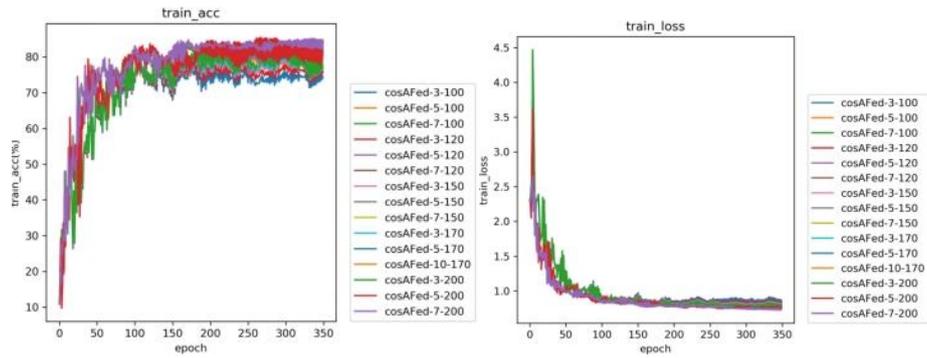

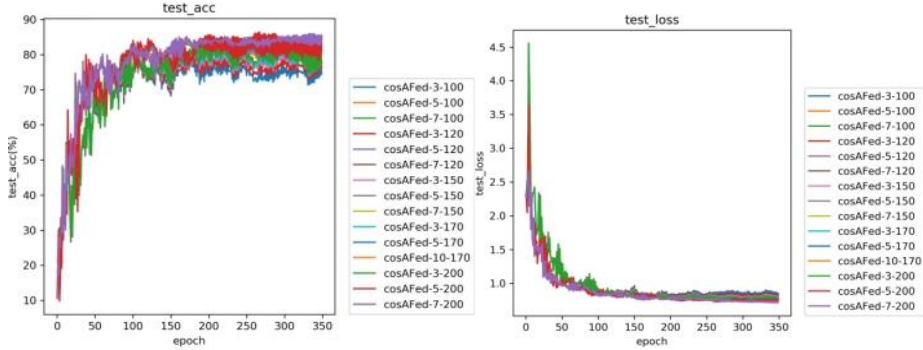

**Fig. 3.** The training results for the Random group with different numbers of selected clients

The training curves for the Equal and Random groups, as shown, indicate that both groups exhibit an upward trend in accuracy curves, with minimal fluctuations towards the end. Similarly, the loss curves for both groups display a downward trend, stabilizing within a narrow range. This observation suggests that during the training process, the accuracy and loss curves change as expected and eventually stabilize, indicating model convergence and the feasibility of the proposed method.

Referring to Figures 2 and 3, it can be observed that, under the same initial privacy budget, increasing the number of selected clients results in reduced fluctuations in the accuracy curve after multiple training rounds. This is because the increase in the number of selected clients leads to a larger dataset used in each training round, making the training process more stable and consequently reducing the fluctuations in the accuracy curve. Additionally, as shown in Figures 4 and 5, increasing the number of selected clients generally leads to an upward trend in best_acc. This is due to two factors: first, the increase in dataset size per training round enhances model accuracy; second, with all other conditions being equal, a higher number of selected clients increases the adjustment factor, thereby increasing the adjusted privacy budget, reducing the added noise, and improving data usability, which in turn enhances model accuracy.

There are instances where an increase in the number of selected clients leads to a decline in best_acc. This is because the increase in the number of clients also results in more noise being added. While the increase in dataset size generally improves model accuracy, it may not fully offset the negative impact of the added noise, causing a slight decline in best_acc. However, this decline is minimal. From the accuracy curves, it is evident that after multiple training rounds, a higher number of selected clients leads to more stable accuracy curves with minimal fluctuations, thus aligning with the observation that the model accuracy curve becomes more stable with an increase in the number of selected clients.

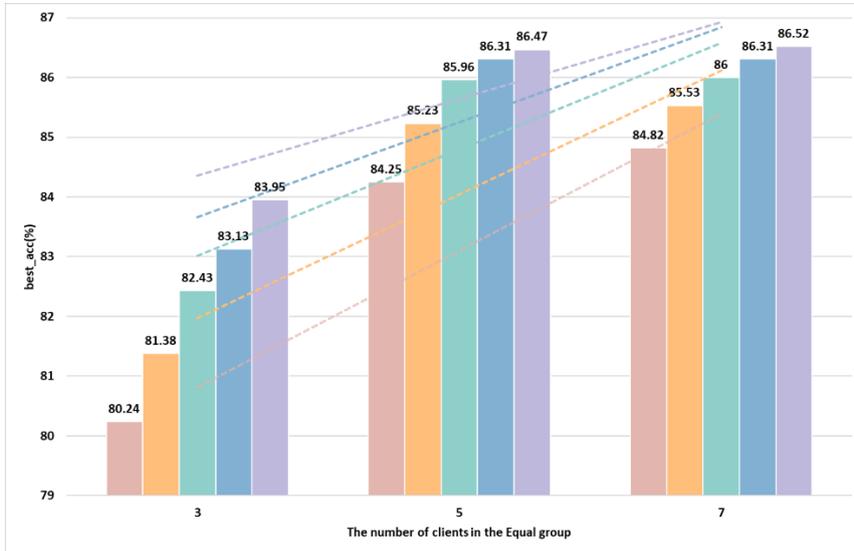

**Fig. 4.** The best_acc of the Equal group with different numbers of selected clients

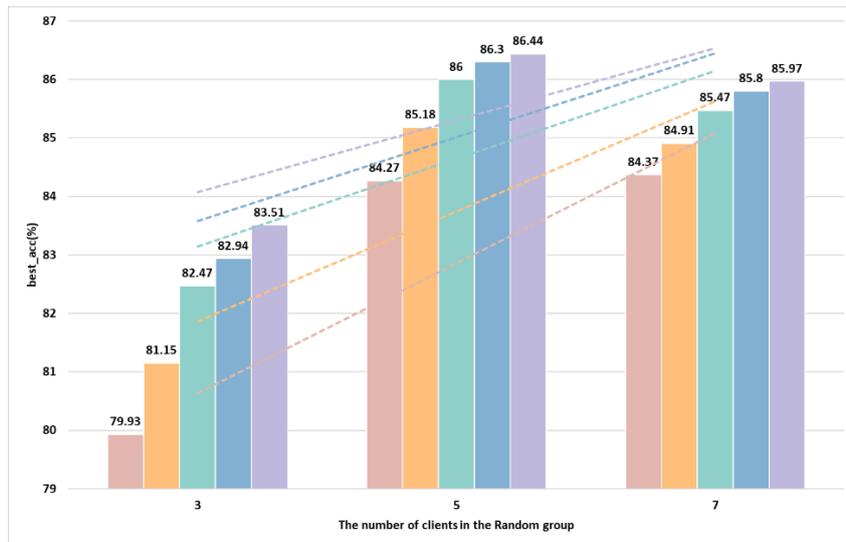

**Fig. 5.** The best_acc of the Random group with different numbers of selected clients

### 5.2.2 Privacy budget

The initial privacy budget for each client in each round is set to 100, 120, 150, 170, and 200, respectively. The number of selected clients in each round is 3, 5, and 7. The experimental results for the Equal group are shown in Figure 6, and the results for the Random group are shown in Figure 7.

The experimental results show that the privacy budget will revert to the initial privacy budget after it is reduced for a while. This is because constantly adding a lot of noise will affect the accuracy and loss of scoring results, resulting in a decrease in the total score. In other words, if a large amount of noise is added, it will negatively affect model training, so the initial privacy budget will be restored to improve the availability of data and thus improve accuracy. In addition, according to the analysis of formula 1, the larger the initial privacy budget, the greater the range of changes in the privacy budget, because when the adjustment coefficient is unchanged, the larger the initial privacy budget, the smaller the calculated privacy budget, and the greater the range of changes. The greater the number of clients selected, the smaller the range of change in the privacy budget. This is because the bigger the number of clients selected, the greater the adjustment coefficient. When the initial privacy budget is unchanged, the larger the calculated privacy budget, the smaller the range of change.

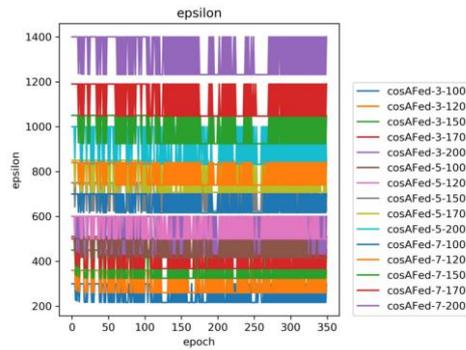

**Fig. 6.** The privacy budget results for the Equal group with different numbers of clients

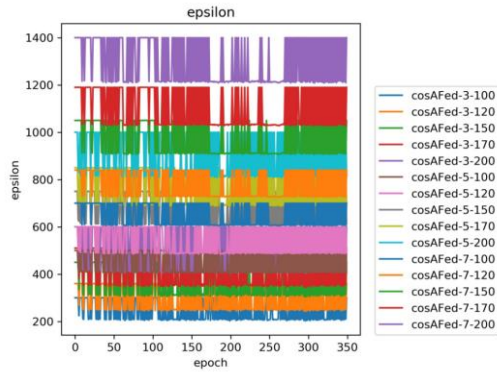

**Fig. 7.** The privacy budget results for the Random group with different numbers of clients

### 5.3    Comparison experiment

In this section, we conduct comparative experiments to validate the superiority of cosAFed over LAPFed, ADPFL [13], and cosFed [11] under the aforementioned experimental conditions. The experimental results are presented in Table 3, Figures 8, and Figure 9. The total privacy budget refers to the sum of the privacy budgets allocated to 3 clients in each round over 350 rounds.

Table 3. The comparative experimental results of cosAFed

| Group | Method | Number of selected clients | Number of rounds | Total privacy budget | best_acc（%） |
|---|---|---|---|---|---|
| Equal | LAPFed | 3 | 350 | 105000 | 81.95 |
|  | ADPFL |  |  | 103155 | 81.86 |
|  | cosFed |  |  | 113853.81 | 82.20 |
|  | cosAFed |  |  | 90016.83 | 80.24 |
| Random | LAPFed | 3 | 350 | 105000 | 80.07 |
|  | ADPFL |  |  | 103170 | 80.06 |
|  | cosFed |  |  | 113829.36 | 80.23 |
|  | cosAFed |  |  | 88372.47 | 79.93 |

The aggregation model obtained by the server uses a fixed initial privacy budget of 100, which remains unchanged during each iteration. This method's constant privacy budget increases the risk of privacy leakage with multiple training rounds, and as the number of rounds increases, the gradient decreases. Excessive noise can reduce model accuracy or even prevent convergence. We made comparisons with other methods based on this foundation.

ADPFL designed a scoring function that lowers the privacy budget based on the results of gradient size, training loss, model accuracy, and the number of communication rounds. This method reduces noise when the gradient becomes smaller and model accuracy is difficult to improve, but the parameter p for reducing the privacy budget is a fixed constant between 0 and 1, making it challenging to set an appropriate value. Because the privacy budget is reduced, the total privacy budget is less than that of LAPFed. Due to the smaller privacy budget and consequently more added noise, the model accuracy decreases in both the Equal and Random groups.

cosFed calculates the cosine value of adjacent global models as a measure of similarity. The privacy budget is adjusted based on this cosine similarity, using a larger privacy budget at the beginning to accelerate model convergence and gradually reducing the budget to protect the model as it nears convergence. Although this approach results in a decreasing trend for the privacy budget, the calculated values are generally higher than the initial privacy budget. Therefore, the total privacy budget is higher than that of LAPFed, and due to less added noise, model accuracy increases in both the Equal and Random groups, though its data protection capability is weaker.

cosAFed builds upon ADPFL and cosFed by considering the cosine value between the previous global model and the current local model, the total number of clients, the number of selected clients, the number of client datasets, and the total number of datasets. It designs an algorithm to adjust the privacy budget and scales the local model updates before adding noise.

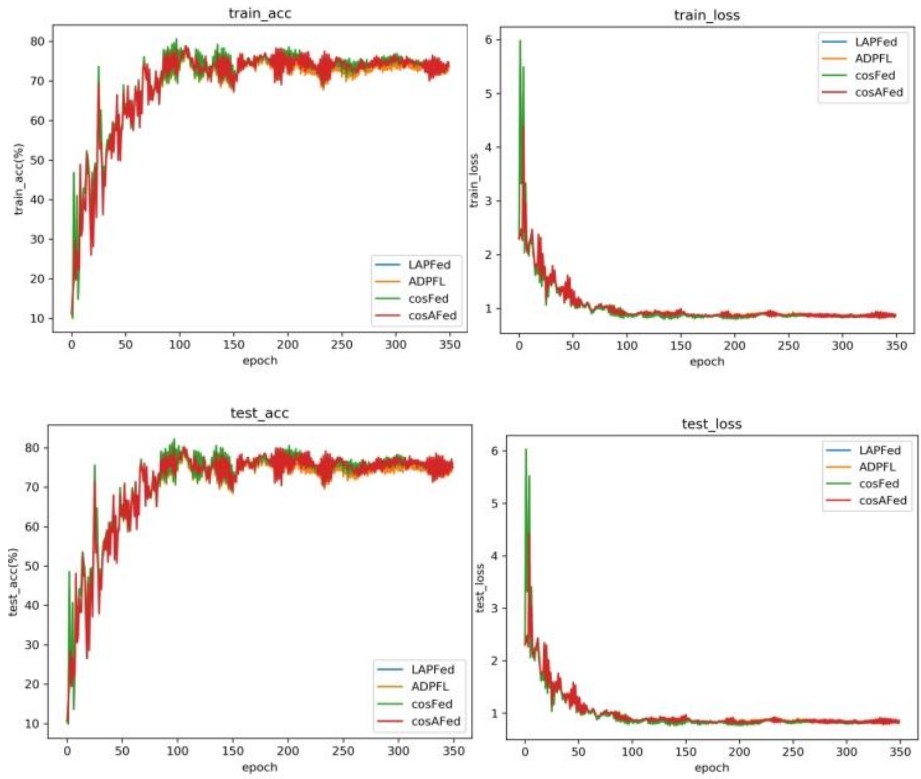

**Fig. 8.** The comparative experimental results of the Equal group for cosAFed

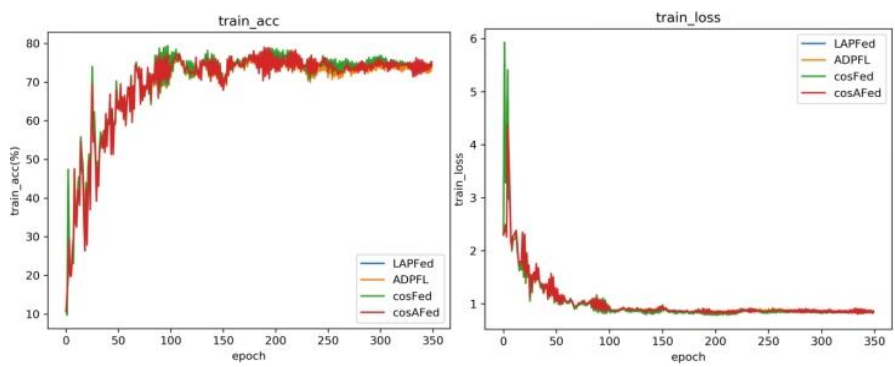

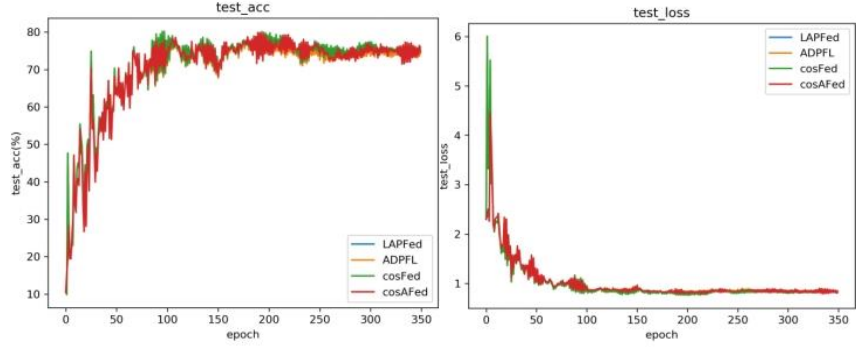

**Fig. 9.** The comparative experimental results of the Random group for cosAFed

Comparative analysis of Table 3, Figures 8, and 9 reveals that, regardless of the Equal or Random group, cosAFed requires the least total privacy budget. Specifically, in the Equal group, cosAFed uses approximately 14% less than LAPFed, 13% less than ADPFL, and 21% less than cosFed. In the Random group, cosAFed uses approximately 16% less than LAPFed, 14% less than ADPFL, and 22% less than cosFed. Therefore, cosAFed has the strongest data protection capability. The specific calculation formula is shown in Formula 9.

$$Relative\ Difference\ Ratio = \left|\frac{Total\ Privacy\ Budget\ of\ cosAFed - Total\ Privacy\ Budget\ of\ LAPFed/ADPFL/cosFed}{Total\ Privacy\ Budget\ of\ LAPFed/ADPFL/cosFed}\right| \quad (9)$$

Additionally, although the model accuracy of cosAFed is slightly lower than that of other methods, the decrease is minimal. In the Equal group, cosAFed's model accuracy is only 1.96% lower than the highest accuracy achieved by cosFed. In the Random group, due to the algorithm's consideration of the dataset proportion in adjusting the privacy budget, the adjustment is more reasonable, resulting in cosAFed's model accuracy being only 0.3% lower than the highest accuracy achieved by cosFed. Therefore, the comparative evaluation shows that in the Equal group, cosAFed has a total privacy budget of 90016.83 and a model accuracy of 80.24%. In the Random group, cosAFed has a total privacy budget of 88372.47 and a model accuracy of 79.93%. This method not only enhances data protection but also ensures model accuracy, demonstrating superiority over other methods.

## 6   Conclusion

With the wide application of federated learning, the problem of privacy protection has also aroused people's attention. As one of the common solutions, differential privacy [19] still needs to consider the balance between security and accuracy, as well as the adaptive addition of noise according to multiple factors. Therefore, we proposed an adaptive differential privacy method based on federated learning. By setting the adjustment coefficient and scoring function, the privacy budget can be dynamically adjusted according to multiple factors. Then, the scaling factor is calculated to process the local model update, and finally adds the noise generated to get the noised local model update. Through method analysis and experimental evaluation, we analyzed the range of parameters and privacy. And we discovered that it reduces the privacy budget by about 16%, while the accuracy remains roughly the same. This method can enhance the data protection ability and ensure the accuracy of the model. In the future, we will further design a more suitable calculation method and use other implementation mechanisms such as Gaussian mechanism to verify the feasibility of this method.